\documentclass[a4paper,twocolumn,citeautoscript,prl]{revtex4-1}

\usepackage{amsmath,bm}
\usepackage[dvips]{graphicx,color}

\setcitestyle{super}

\begin{document}

\title{Ultra-weak long-range interactions of solitons observed over astronomical distances}

\author{Jae K. Jang}
\author{Miro Erkintalo}
\author{Stuart G. Murdoch}
\author{St\'ephane Coen}

\affiliation{Department of Physics, The University of Auckland, Private Bag 92019, Auckland 1142, New Zealand}

\begin{abstract}
  \noindent We report what we believe is the weakest interaction between solitons ever observed. Our experiment
  involves temporal optical cavity solitons recirculating in a coherently-driven passive optical fibre ring
  resonator. We observe two solitons, separated by up to 8,000~times their width, changing their temporal
  separation by a fraction of an attosecond per round-trip of the 100~m-long resonator, or equivalently 1/10,000
  of the wavelength of the soliton carrier wave per characteristic dispersive length. The interactions are so weak
  that, at the speed of light, they require an effective propagation distance of the order of an astronomical unit
  to fully develop, i.e.\ tens of millions of kilometres. The interaction is mediated by transverse acoustic waves
  generated in the optical fibre by the propagating solitons through electrostriction.
\end{abstract}

\maketitle

\noindent Solitons are self-localized wave packets that do not spread, the dispersion of the supporting medium being
cancelled by a nonlinear effect \cite{russell_report_1845, zabusky_interaction_1965, hasegawa_transmission_1973,
akhmediev_solitons_1997}. They are universal, and in many respects behave like particles
\cite{zabusky_interaction_1965}. They exert forces on each other and can interact in various ways, elastically or
inelastically \cite{zabusky_interaction_1965, stegeman_optical_1999, craig_solitary_2006}. Here we report what we
believe is by far the weakest form of soliton interaction ever observed. Using recirculating optical cavity
solitons, we report interactions so weak that the solitons shift their positions by only about $10^{-7}$ of their
width --- amounting to 1/10,000 of the wavelength of the soliton carrier wave --- per characteristic dispersive
length. At the speed of light, these interactions require effective propagation distances of the order of an
astronomical unit (AU) to be revealed, i.e.\ tens of millions of kilometres. The sheer fact that we can actually
observe such ultra-weak interactions in a noisy laboratory environment highlights the robustness and stability of
solitons as never-before.

Solitons occur in media as diverse as water, DNA, plasma, or ultra-cold gases, \cite{russell_report_1845,
gardner_method_1967, lonngren_soliton_1983, saha_long-range_2012, polturak_solitonlike_1981, burger_dark_1999,
denschlag_generating_2000} but over the last 20 years optics has led the way in our understanding of soliton
interactions because of the ease with which optical solitons can be studied experimentally \cite{bjorkholm_cw_1974,
mollenauer_experimental_1980, barthelemy_propagation_1985, segev_optical_1998}. Optical solitons have been shown to
attract, repel, breakup, merge, orbit each-other or even annihilate \cite{gordon_interaction_1983,
reynaud_optically_1990, tikhonenko_three_1996, shih_three-dimensional_1997, krolikowski_annihilation_1998,
stegeman_optical_1999}. As in various other media, soliton interactions can be either short-range, occurring when
the tails of neighbouring solitons overlap \cite{gordon_interaction_1983, snyder_self-induced_1991,
mitschke_soliton_1998, craig_solitary_2006}, or long range, through a coupling with a non-local response, be it
optical radiation, charge carriers, or thermal waves \cite{smith_experimental_1989, rotschild_long-range_2006,
skryabin_colloquium_2010, allen_long-range_2011}. The weakest soliton interactions reported are long range
\cite{smith_experimental_1989, rotschild_long-range_2006}, but their observation is typically limited by the
duration or propagation length over which the solitons can be maintained. In optics, this is often simply dictated
by the size of a nonlinear crystal or the length of an optical fibre. To overcome this restriction, our experiment
involves solitons recirculating in an optical fibre loop.

More specifically, we consider temporal cavity solitons (CSs) propagating in a coherently-driven passive nonlinear
fibre ring resonator \cite{leo_temporal_2010}. Not only are these objects genuine solitons, for which the nonlinear
self-trapping occurs in the longitudinal (temporal) direction, but the losses they suffer each round-trip in the
fibre loop are also compensated, thanks to the continuous-wave (c.w.)\ beam coherently driving the system
\cite{tlidi_localized_1994, firth_cavity_2002, barland_cavity_2002, lugiato_introduction_2003}. In this respect, CSs
are dissipative solitons \cite{akhmediev_dissipative_2008}, and can persist indefinitely in the driven resonator. In
our fibre experiment, we find that temporal CSs interact over long ranges through sound. In this scenario, the
electric field of a first CS deforms the fibre material --- in a process known as electrostriction --- and excites
an acoustic wave that propagates outwards across the fibre core and cladding. The associated density variation leads
to a small time-varying change of the refractive index. When passing through this perturbation, a second CS
undergoes a slight shift of its carrier frequency, which --- due to dispersion --- changes its group velocity
\cite{dianov_long-range_1992}. In this way, the trailing CS either catches up with the leading one (attraction) or
is delayed (repulsion). Such acoustic interactions have been studied in the past in the context of optical fibre
telecommunication systems with the traditional Kerr solitons of the Nonlinear Schr\"odinger (NLS) equation
\cite{smith_experimental_1989, dianov_long-range_1992, townsend_measurement_1996, jaouen_transverse_2001}. With
temporal CSs, an important difference arises however because they are robust attracting states of the nonlinear
resonator, tied to the phase and carrier frequency of the coherent driving field \cite{firth_temporal_2010}. So
while NLS solitons can shift their carrier frequency, and hence their velocity, at will, CSs are severely limited to
do so. As a matter of fact, our study reveals that the strength of the acoustic interactions of CSs is about
10,000--100,000 times weaker than in single-pass configurations.

\section*{Setup}

\begin{figure*}[t]
  \centerline{\includegraphics{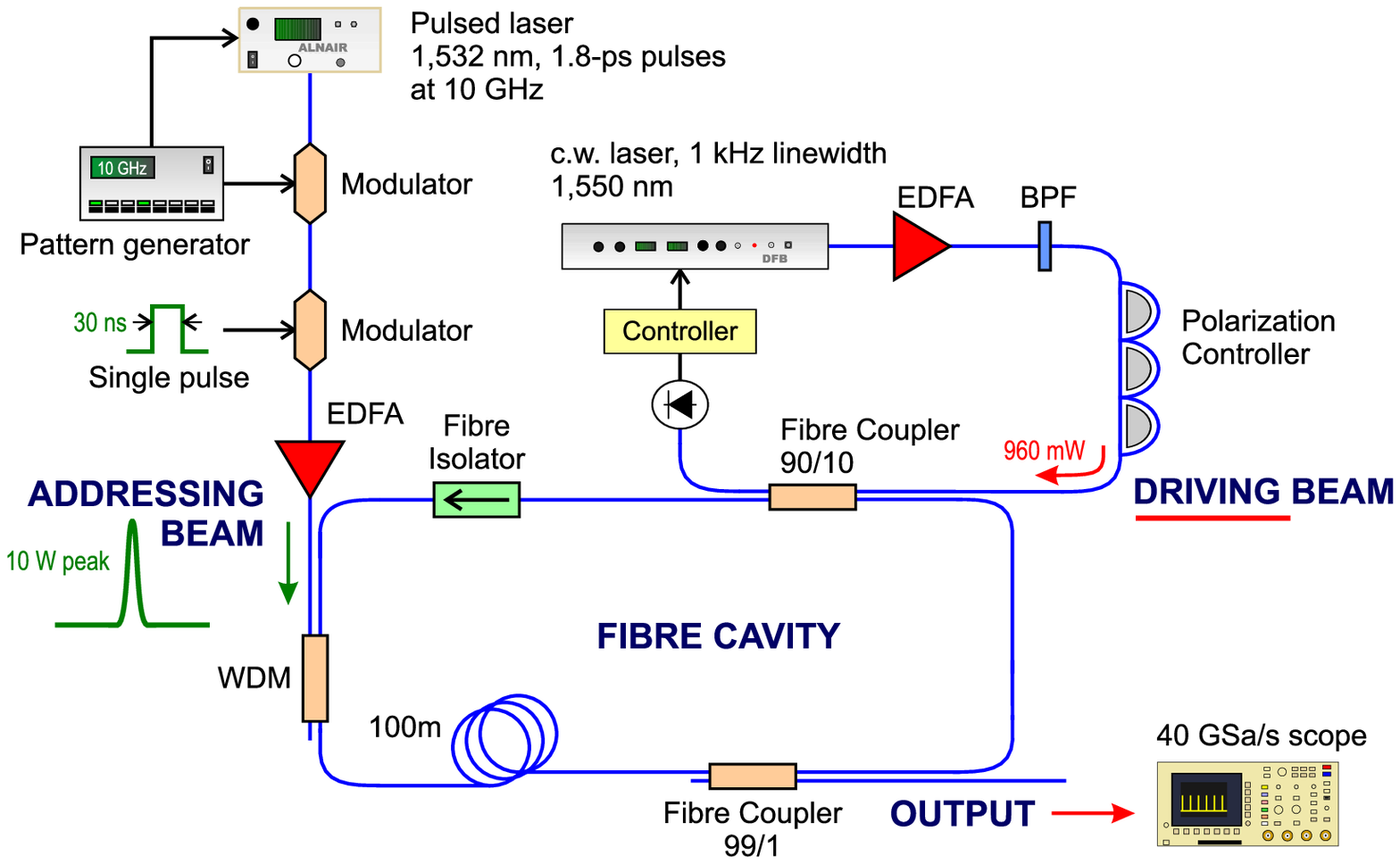}}
  \caption{\textbf{Experimental setup. } A passive fibre cavity is externally driven by a c.w.\ laser beam. The part
    of this beam that is reflected off the cavity is used to actively lock the driving laser frequency to the cavity
    resonance. Temporal cavity solitons are excited with $1.8$-ps addressing pulses at a different wavelength and
    coupled into the cavity through a wavelength-division multiplexer (WDM). The number of solitons excited and their
    separation is set with two intensity modulators and a pattern generator. Erbium-doped fibre amplifiers (EDFA)
    boost the power of the two beams while optical bandpass filters (BPF) improve the signal-to-noise ratio of the
    measurements. The output is analysed with an ultra-fast real time oscilloscope.}
  \label{fig:setup}
\end{figure*}
\noindent Our experimental setup is depicted in Fig.~\ref{fig:setup}. The passive fibre ring resonator is about
100~m long and entirely made of standard silica telecommunication fibre closed on itself by a 90/10 fibre coupler.
It takes light approximately $0.5\ \mu$s to complete a single round-trip of the cavity. The coupler is arranged so
that 90\,\% of the intracavity power is recirculated, which leads to a large effective nonlinearity. Solitons occur
due to the intensity-dependent refractive index of silica (the Kerr effect), which has a nearly instantaneous
response \cite{hasegawa_transmission_1973, agrawal_nonlinear_2006}. The resonator incorporates an optical fibre
isolator to prevent the build-up of stimulated Brillouin scattering radiation, which would otherwise deplete the
driving beam \cite{agrawal_nonlinear_2006}, a wavelength-division multiplexer (WDM) to couple the addressing pulses
used to excite CSs, and a 1\,\% output coupler to monitor the intracavity dynamics. Overall, the total round-trip
losses are about 29\,\%, corresponding to a cavity finesse of 22 and to 91-kHz-wide resonances. The cavity is
coherently driven by a c.w.\ laser running at 1,550~nm wavelength, with a narrow linewidth of 1~kHz. It is boosted
by an Erbium-doped fibre amplifier before being launched into the cavity. The part of the driving beam that is
reflected off the cavity is directed to a servo-controller, which controls the laser frequency. In this way, the
driving laser is actively locked on a cavity resonance. In comparison to Leo et~al who were only able to maintain
the CSs for a few seconds \cite{leo_temporal_2010}, our locking system is significantly more robust and works for a
larger excursion of the environmental parameters. Combined with the fact that our cavity is about four times
shorter, we can routinely achieve stable locking of the cavity for periods in excess of 30~minutes. This is a key
factor in observing the weak interactions we report here.

\begin{figure*}[t]
  \centering
  \includegraphics[]{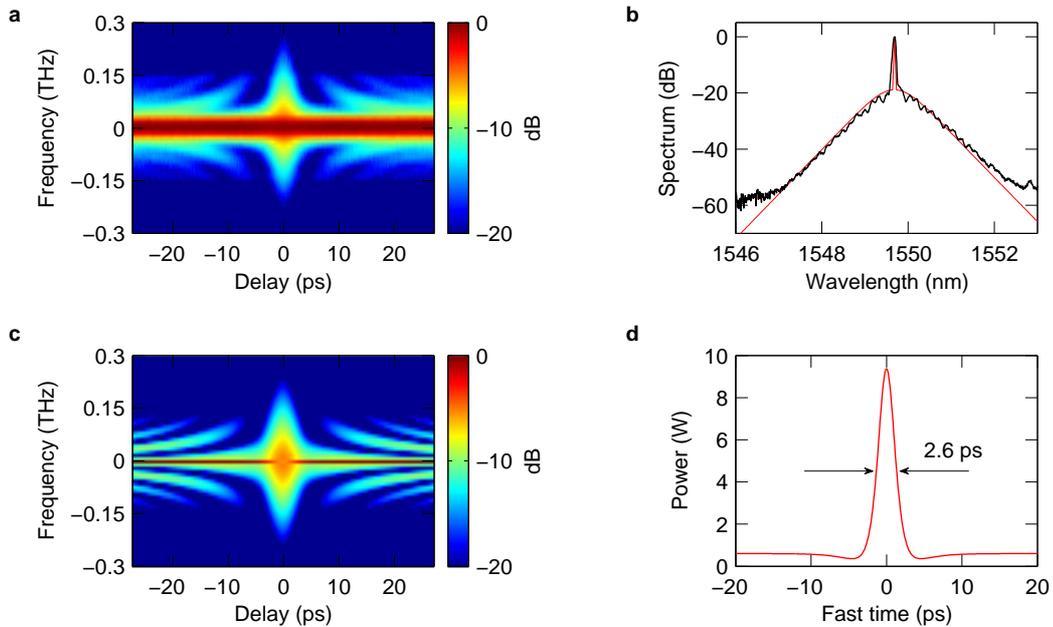}
  \caption{\textbf{Cavity soliton characteristics.}
    \textbf{a,} Experimental FROG trace and \textbf{b,} optical spectrum (black) of CSs. \textbf{c},
    Simulated FROG trace in excellent agreement with \textbf{a}. \textbf{d,} Corresponding
    numerical temporal intensity profile of the CSs, which reveals their $2.6$~ps duration. The simulated spectrum is
    shown in red in~\textbf{b}. The horizontal line and the hyperbolic fringes visible at and around the middle of
    the FROG trace, as well as the peak at the centre of the optical spectrum, are all signatures of the
    c.w.\ background on which the CSs are superimposed, and which is clearly visible in~\textbf{d}.}
  \label{fig:frog}
\end{figure*}
The CSs are excited using the same technique as in Ref.~\citenum{leo_temporal_2010}, i.e.\ through cross-phase
modulation between the intracavity c.w.\ background field and ultra-short pulses at a different wavelength. These
addressing pulses are $1.8$~ps in duration and are generated by a pulsed laser with a repetition rate locked to the
10~GHz clock of an electronic pattern generator (see left part of Fig.~\ref{fig:setup}). A series of two intensity
modulators is used to select individual addressing pulses so as to control the number of CSs excited into the
cavity, as well as their initial separation (in multiples of the 100~ps period of the 10~GHz pulsed laser). The
first modulator imprints the repetitive bit pattern of the generator onto the laser pulse train. The resulting
optical signal is then gated by the 2nd modulator so as to select a single period of that pattern when starting
measurements. The rest of the time, the 2nd modulator is completely blocking the addressing beam. Note that the CSs
will have the same wavelength as that of the driving beam \cite{firth_cavity_2002,leo_temporal_2010}.

Once excited, temporal CSs circulate stably in the fibre resonator, experiencing no changes in their shape or
amplitude. This confirms their solitonic nature. We characterized them by measuring a time-frequency representation
of their optical field using FROG (frequency resolved optical gating) \cite{trebino_measuring_1997} as well as their
optical spectrum (Fig.~\ref{fig:frog}a,b). The central structure of the FROG trace signals the presence of short
pulses of a few picoseconds duration circulating in the cavity. The FROG trace is also barred in the centre by an
horizontal line. This line signals the presence of a c.w.\ background on which the pulses are superimposed. Such
c.w.\ background is a central characteristic of CSs and is expected \cite{firth_cavity_2002, barland_cavity_2002,
leo_temporal_2010}. The central (DC) peak of the optical spectrum is another signature of the background as are the
spectral interferometric fringes visible around the main peak of the FROG trace. To understand the latter, it is
worth pointing out that the FROG measurement is performed by spectrally resolving, as a function of delay, a signal
consisting of the optical field under study multiplied by its own delayed copy \cite{trebino_measuring_1997}. In
presence of a pulse on a c.w.\ background, and for delays larger than the pulse duration, the FROG signal will be
made up of two copies of the pulse, naturally leading to fringes in the spectral domain, and we have successfully
verified that the observed fringes follow hyperbolic trajectories in the time-frequency plane $(\tau,f)$,
$f=1/\tau$, $2/\tau$, $3/\tau$, $\ldots$ The presence of these fringes therefore also highlights the coherence of
the CS with respect to the c.w.\ background, and hence with the driving field. All these features are also clearly
visible on a numerically simulated FROG trace (Fig.~\ref{fig:frog}c), with the corresponding temporal intensity
profile shown in Fig.~\ref{fig:frog}d. The excellent agreement between the experimental and simulated FROG traces
allows us to infer that our CSs are about $2.6$~ps (full-width at half maximum) in duration (Fig.~\ref{fig:frog}d).
Note that the experimental FROG trace is somewhat blurred in comparison to the numerical one due to the limited
spectral resolution of the measuring apparatus.

\section*{Results}

\begin{figure*}[t]
  \centering
  \includegraphics[width=15cm]{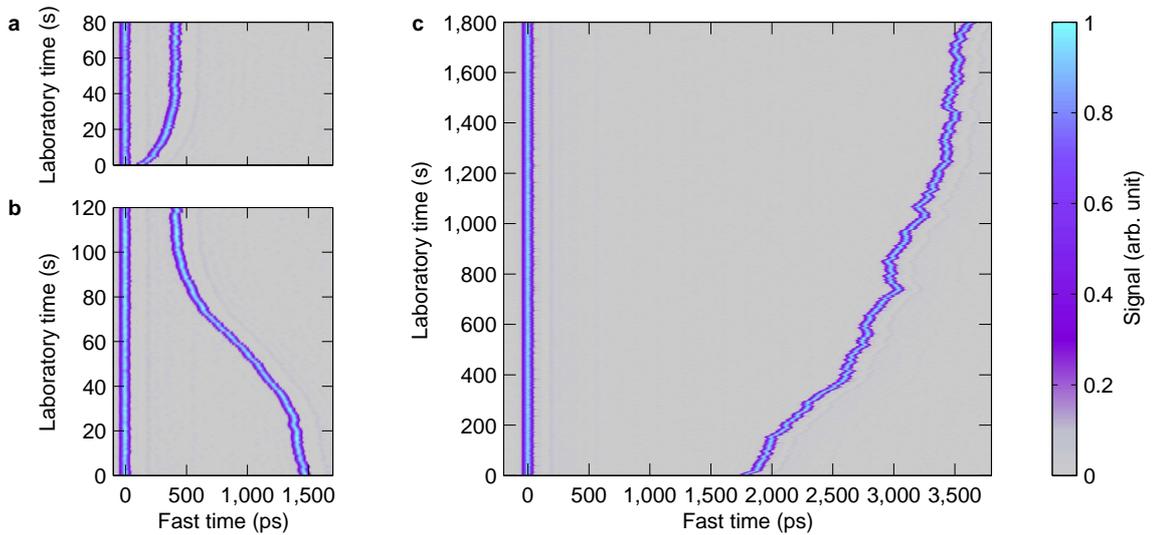}
  \caption{\textbf{First example of long-range interactions of a pair of CSs.} Colour plots made up of successive
    oscilloscope measurements at the cavity output showing the temporal evolution of two CSs. \textbf{a,} For an
    initial separation of 100~ps (38 soliton widths), the interaction is repulsive until a stable separation of 420~ps
    is attained. \textbf{b,} An attractive interaction is observed when the initial separation is increased to
    1,500 ps (577 soliton widths), eventually leading to the same final separation of 420~ps. \textbf{c},~For
    an initial separation of 1,800~ps, the interaction is very weakly repulsive. The colour map applies to all
    three panels.}
  \label{fig:400ps}
\end{figure*}
\noindent The interactions between the CSs were studied by writing a single pair of CSs into the cavity and by
observing through the output coupler how the temporal separation between the two pulses evolves over many
round-trips. Measurements were performed with an ultra-high-sampling-rate real-time digital oscilloscope triggered
on the leading pulse of the pair. In Fig.~\ref{fig:400ps}a, we show a colour plot of successive oscilloscope traces
when the two CSs are initially separated by 100~ps, more than 38~soliton widths. Despite the large separation, the
trailing CS can be seen to be gradually repelled away from the leading one. This long-range repulsion persists for
about 20~seconds until a stationary separation of 420~ps is reached (corresponding to about 160 soliton widths).
Note that the oscilloscope has a temporal resolution of about 50~ps, significantly larger than the $2.6$~ps CS
duration, hence the plot does not do full justice to the large difference of scales between the width of the CSs and
their separation. In Fig.~\ref{fig:400ps}b, the initial separation has been set to 1,500~ps (or 577~soliton widths).
The interaction is now attractive, and again persists until the same stationary separation of 420~ps is reached. We
find that beyond an initial separation of 1,500~ps, the interaction becomes repulsive again. Fig.~\ref{fig:400ps}c
illustrates this latter case, where starting from a separation of 1,800~ps (or 692~soliton widths), the trailing
pulse slowly moves away from the leading one, to reach a separation of about 3,600~ps after 30~minutes. It is clear
that the repulsion observed in Fig.~\ref{fig:400ps}c is significantly weaker than in Fig.~\ref{fig:400ps}a.

The soliton interactions reported above take place over timescales which are incredibly slow for an optical system.
The differences in orders of magnitude can be appreciated by considering that during the 30~minute interaction shown
in Fig.~\ref{fig:400ps}c, the CSs travel around the cavity $3.6$~billion times, for a total distance of 360~million
kilometres, or $2.4$~AU, i.e.\ more than twice the distance between the Earth and the Sun. Yet over that vast
distance, the two CSs only shift their relative position by less than 2~ns, equivalent to a change of 40~cm in
spatial separation, 12~orders of magnitude smaller than the distance travelled. Such displacement corresponds to
half an attosecond per cavity round-trip or equivalently to $2\times 10^{-7}$ soliton widths, or 1/10,000 of the
driving laser wavelength, per linear dispersive length (the distance $L_\mathrm{D}$ over which a Gaussian pulse of
the same duration would broaden by a factor of $\sqrt{2}$ due to linear dispersion only
\cite{agrawal_nonlinear_2006}; in our case $L_\mathrm{D}=114$~m).

\begin{figure*}[t]
  \centering
  \includegraphics[clip,width=14.2cm]{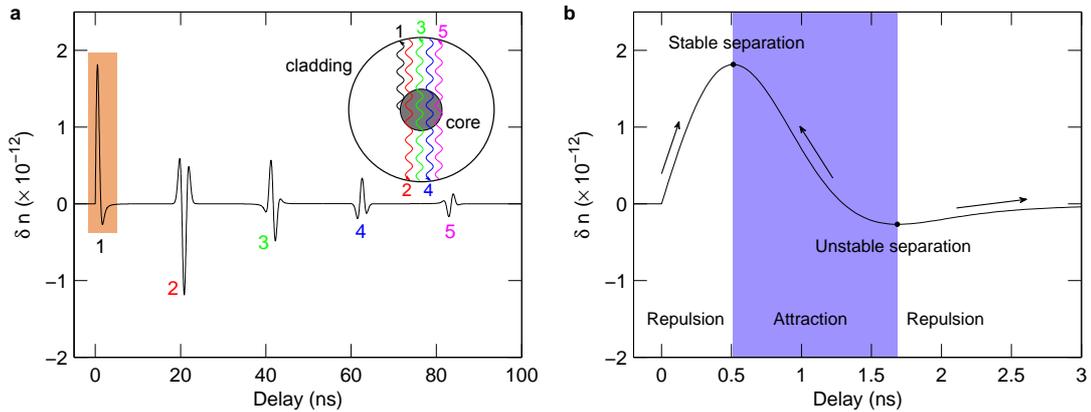}
  \caption{\textbf{Acoustic response.} \textbf{a,} Theoretical impulse response of the refractive index acoustic
    perturbation calculated in our experimental conditions, and scaled to the CS energy. Subsequent spikes
    (numbered 1--5) are separated by about 21~ns and arise from consecutive reflections of the acoustic wave
    from the fibre cladding-coating boundary as schematically illustrated in the inset. \textbf{b,} Close-up on the
    first peak of the acoustic response. Ranges of repulsion and attraction of a trailing CS are highlighted. The
    maximum corresponds to a stable separation.}
  \label{fig:acoustic}
\end{figure*}
It is worth noting that the interaction strength between solitons propagating in an ideal instantaneous Kerr medium
decays exponentially if they are separated by more than a few soliton widths \cite{gordon_interaction_1983,
mitschke_experimental_1987, reynaud_optically_1990}. To understand the physics of the long-range interactions
illustrated in Fig.~\ref{fig:400ps}, we have to consider the time-varying refractive index perturbation mediated by
an acoustic wave excited by the leading CS. The impulse response of that perturbation can be calculated using the
method of Ref.~\citenum{dianov_long-range_1992} and is plotted in Fig.~\ref{fig:acoustic}a, scaled to the energy of
our CSs. We must first point out the smallness of this refractive index perturbation. With a maximum of
approximately one part-per-trillion, it is more than three orders of magnitude smaller than the Kerr-induced index
change due to the peak power of the CS. Also, as can be seen in Fig.~\ref{fig:acoustic}a, the impulse response
consists of short spikes (1--2~ns in duration) occurring every 21~ns. These correspond to increasing orders of
echoes, i.e.\ back-and-forth reflections of the acoustic wave from the fibre cladding-coating boundary back into the
fibre core region \cite{jaouen_transverse_2001}. Given the standard cladding diameter of $125\ \mu$m of our fibre,
21~ns perfectly matches with the travel time of sound in silica at 5,996~$\mathrm{m/s}$
\cite{fellegara_measurement_1997}. In between the spikes, the response remains negligible due to the absence of
overlap between the optical mode trapped inside the fibre core and the acoustic wave travelling in the cladding.

The time-varying nature of the refractive index perturbation $\delta n$ is responsible for a shift in the
instantaneous frequency of the trailing CS. The slope of the perturbation at the delay of the trailing CS determines
whether the interaction is repulsive or attractive. If $d\delta n/dt>0$ (respectively, $<0$), the trailing CS is
red-shifted (blue-shifted) with respect to the leading one. Because our experiment is performed in the anomalous
dispersion regime \cite{agrawal_nonlinear_2006}, this frequency shift translates into a smaller (larger) group
velocity, resulting in an effective repulsion (attraction) between the CSs. It is only when $d\delta n/dt=0$ that
the two CSs travel at the same group velocity and do not move with respect to each other. A close-up of the first
spike of the acoustic impulse response plotted in Fig.~\ref{fig:acoustic}a is shown in Fig.~\ref{fig:acoustic}b
where we have highlighted, based on the above analysis, the ranges of repulsion and attraction. Furthermore, from
this figure, it is straightforward to verify that the maximum of the refractive index perturbation corresponds to a
stable separation while the minimum is unstable. This predicted behaviour is in remarkable qualitative agreement
with the observations described in Fig.~\ref{fig:400ps}. In particular, the weakness of the repulsion observed for a
separation larger than about 1,500~ps simply results from the small (positive) slope of the refractive index
perturbation in that region and all the way to the first echo at 21~ns. We must note that the measured stable
separation of 420~ps does not quite match with the 510~ps at which the theoretical maximum occurs (see
Fig.~\ref{fig:acoustic}b). However, the transverse acoustic model of optical fibres is known to be deficient and has
systematically failed in precisely reproducing the experimental measurements of the first acoustic contribution
\cite{townsend_measurement_1996, buckland_mode-profile_1999, biryukov_excitation_2002}. Incidentally, our results
may constitute the most accurate measurement of this primary refractive index extremum to date.

\begin{figure*}[t]
  \centering
  \includegraphics[width=14cm,clip]{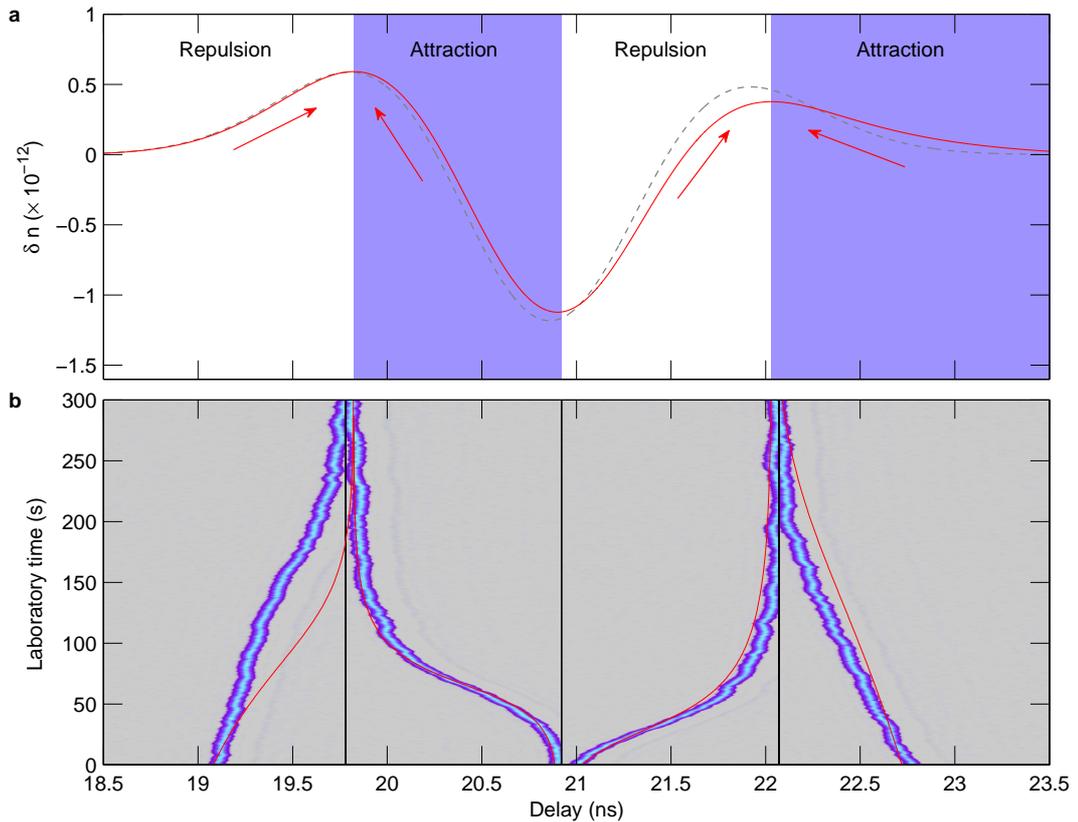}
  \caption{\textbf{Interactions of two CSs mediated by the first acoustic echo.} \textbf{a,} Close-up of
    the first echo of the acoustic impulse response (dashed; see spike labelled~2 in Fig.~\ref{fig:acoustic}a). The red
    curve shows the actual index change when taking into account the exact CS intensity profile and its perturbed
    background. Two stable separations of CSs exist in the range shown. \textbf{b,} Experimental colour plot
    of the trailing CS in each of the region highlighted in \textbf{a} (as in Fig.~\ref{fig:400ps},
    using the same colour map, but with the leading CS omitted for clarity). The plot consists of the juxtaposition
    of four different independent measurements (separated by vertical lines) obtained for different initial
    separations between the two CSs. Red curves are numerical simulations.}
  \label{fig:20ns}
\end{figure*}
The acoustic origin of the interaction can be confirmed by probing the dynamics of a pair of CSs whose initial
separation is adjusted around 21~ns to match the re-entrance of the first acoustic echo into the fibre core. Note
that this separation is still much smaller than the round-trip time of $0.5\ \mu$s so that the CS pair can truly be
considered as isolated. A close-up of the impulse response of the refractive index perturbation about the first echo
is plotted in Fig.~\ref{fig:20ns}a (dashed curve). The response here has two maxima, and for this range of
separations we would therefore expect to observe two different stable separations. Experimental measurements are
juxtaposed on the same graph in Fig.~\ref{fig:20ns}b for four different initial separations sampling the four
regions of repulsion and attraction around 21~ns separation. Note that here we have omitted to plot the leading CS
(at zero delay) for clarity (at the scale of Fig.~\ref{fig:20ns}b, including the leading CS would make the figure
three pages wide). We observe stable separations of $19.79$~ns and $22.05$~ns. This is in excellent agreement with
the maxima of the actual acoustic index change (red curve in Fig.~\ref{fig:20ns}a) calculated as the convolution of
the impulse response (shown in Fig.~\ref{fig:acoustic}a) with the exact leading CS intensity profile and its
perturbed background (see Methods). Note how the maxima of the impulse response are shifted towards slightly longer
delays when considering the more complete convolution calculation, from $19.80$~ns and $21.92$~ns, to $19.82$~ns and
$22.03$~ns, respectively. For this case, we have also performed a full modelling of the interactions (see Methods)
and the corresponding numerical trajectories of the trailing CS are superimposed on the experimental data as the red
curves in Fig.~\ref{fig:20ns}b. Considering the large difference of timescales involved --- $2.6$~ps CSs shifting
their position by about 1~ns over 100--200 seconds
--- and the absence of any fit parameter, the agreement is spectacularly good. Most of the discrepancies are likely
due to incomplete knowledge of the acoustic impulse response or mechanical and thermal fluctuations in the
experimental environment. These observations and simulations confirm the acoustic nature of the interactions we
observe. We must point out that 21~ns corresponds to more than 8,000~soliton widths. This is by far the longest
interaction range ever reported for solitons.

\section*{Discussion}

\noindent Our study provides clear experimental evidence of record-breaking ultra-weak interactions between
solitons. In our experiment, interacting solitons shift their relative position by an amount as small as $2 \times
10^{-7}$ of their width, or 1/10,000 of the soliton carrier wavelength, per linear dispersive length. This
represents a microscopic displacement ($\sim 100$~pm) over a macroscopic propagation distance ($\sim 100$~m), with
12~orders of magnitude difference. The fact that such a weak interaction can accumulate in a well defined manner
over distances of several hundred million kilometres (or billions of dispersive lengths) is also truly remarkable.
It highlights the extreme stability, robustness, and coherence of the process, and of solitons in general. The
long-range nature of the interaction further compounds these feats. Indeed, our observations are performed with
solitons separated by up to 8,000~times their width. Translated into spatial units, our solitons localized within
$0.5$~mm of fibre interact across a 4~m separation. All together, these figures are simply spectacular.

Our observations make clear that in the context of our study, which is performed with temporal optical cavity
solitons, the interactions are mediated by acoustic waves. We believe our findings explain the repulsion of unknown
origin described in Ref.~\citenum{leo_temporal_2010}, which reported the first experimental realization of temporal
CSs. In that work, the evolution of the separation between pairs of CSs could not be tracked over time because the
cavity was only stable for a few seconds. Clearly, the remarkable stability of our experiment, with the cavity
interferometrically stabilized over periods in excess of 30 minutes, was a key factor for the success of our
measurements. Finally, we must stress that temporal CSs have recently been shown to be the likely underlying
temporal structure of broadband frequency combs generated in high-Q Kerr resonators \cite{coen_modeling_2012}. As
acoustic interactions have been shown to play a role in the stabilization of the repetition rate of harmonically
mode-locked fibre lasers \cite{gray_femtosecond_1995, pilipetskii_acoustic_1995}, it is interesting to speculate
whether they have a similar role in the context of Kerr frequency comb generation and cavity optomechanics
\cite{kippenberg_analysis_2005}.

\section*{Methods}

\footnotesize

\subparagraph*{\hskip-10pt Experimental setup.}

The resonator is made up of 100~m of standard telecommunications single-mode silica optical fibre
(Corning{\tiny\textsuperscript{\textregistered}} SMF-28{\tiny\texttrademark}). The choice of a single-mode waveguide
enables us to make observations unhampered by the diffraction of the beams \cite{agrawal_nonlinear_2006}: The system
is purely one-dimensional. At 1,550~nm wavelength, the fibre presents a second-order dispersion coefficient $\beta_2
= -21.4\ \mathrm{ps^2/km}$ (measured by white-light interferometry) and a c.w.\ nonlinearity coefficient, inclusive
of the electrostrictive contribution, $\gamma \simeq 1.2\ \mathrm{W^{-1}\,km^{-1}}$ (measured from the nonlinear
tilt of the cavity resonance) \cite{agrawal_nonlinear_2006}. The intracavity isolator preventing stimulated
Brillouin scattering is polarization-independent and has a 60-dB isolation factor. The WDM coupler used to inject
the addressing pulses is a filter WDM coupler with a 10-nm wide transition band centred at 1,540~nm. The
free-spectral-range (FSR) and finesse of the resonator are measured to be $2.07$~MHz and~22, respectively. The
driving beam is generated by a Koheras AdjustiK{\tiny\texttrademark{}} E15 laser, which is an Erbium-doped
distributed feedback fibre laser. It presents a linewidth $< 1$~kHz and is operated at a 20~mW output power level.
After amplification, we use a narrow bandpass filter (BPF, $0.5$~nm bandwidth at $0.5$~dB) centred on the driving
laser wavelength to reject most of the amplified spontaneous emission (ASE) noise of the amplifier before the
driving beam is launched into the cavity. The driving power at the input to the resonator is 960~mW. A commercial
100~kHz PID system (SRS SIM960) is used to lock the driving power reflected off the resonator to a set level
allowing for precise control of the cavity detuning. The error signal of the PID controller is directly fed to the
fast piezo-electric tuning system of the AdjustiK{\tiny\texttrademark{}} laser. The addressing pulses are derived
from a tunable semiconductor mode-locked laser (Alnair MLLD-100) centred at 1,532~nm with a pulse width of $1.8$~ps
and are subsequently amplified to a peak power of 10~W. The two intensity modulators used to select individual
addressing pulses have a $12.5$~GHz bandwidth. For the oscilloscope measurements of Fig.~\ref{fig:400ps}a--c
and~\ref{fig:20ns}b, detection was performed with a $12.5$~GHz PIN amplified photodiode. The oscilloscope had a
bandwidth of 12~GHz and a sampling rate of 40~GSample/s, and traces were acquired every second. Also, to improve the
dynamic range of the measurements, the c.w.\ background on which the CSs are superimposed was filtered out by a
narrow BPF ($0.5$~nm bandwidth) centred at 1,551~nm, slightly off the driving laser wavelength, and placed in front
of the photodiode \cite{leo_temporal_2010}.

\subparagraph*{\hskip-10pt FROG measurements.}

The time-frequency representation of Fig.~\ref{fig:frog}a was obtained with a second-harmonic frequency-resolved
optical gating (SHG-FROG) apparatus (Southern Photonics HR150) with an $0.09$~nm (or 45~GHz) spectral resolution at
the second-harmonic wavelength. We must point out that we cannot measure the FROG trace of CSs when there is only a
single CS circulating in the cavity because the energy contained in the CS c.w.\ background is four orders of
magnitude larger than the energy of the actual $2.6$~ps-wide pulse, which is too large for the dynamic range of the
FROG spectrometer. For that reason, the FROG measurement is obtained by filling the resonator with about 1,800~CSs.
As all the CSs are identical, the result is the same as that with a single CS, except for the apparent amplitude of
the background signal at DC in comparison to the wings of the spectrum. The number of CSs circulating in the cavity
was used as a fit parameter to calculate the numerically simulated FROG trace of Fig.~\ref{fig:frog}c. The spectrum
shown in Fig.~\ref{fig:frog}b was measured with an optical spectrum analyser presenting an $0.05$~nm spectral
resolution, in the same conditions as the FROG trace.

\subparagraph*{\hskip-10pt Modelling of the acoustic response.} We list here the parameters used to calculate the
acoustic response of our silica fibre, which is plotted in Fig.~\ref{fig:acoustic}a, based on the model of
Ref.~\citenum{dianov_long-range_1992}. We used $\rho = 2210\ \mathrm{kg/m^3}$ for the density of silica
\cite{melloni_direct_1998}, $v = 5996\ \mathrm{m/s}$ and $v_\mathrm{s} = 3740\ \mathrm{m/s}$ for, respectively, the
longitudinal and transverse (shear) velocity of sound \cite{fellegara_measurement_1997, biryukov_excitation_2002},
$\gamma_\mathrm{e} = \rho d\varepsilon/d\rho = 0.902$ for the electrostrictive constant (where $\varepsilon$ is the
dielectric constant) \cite{melloni_direct_1998}, $\Gamma_\mathrm{n} = 3 \times 10^7\ \mathrm{s^{-1}}$ for the
acoustic damping coefficient \cite{dianov_long-range_1992}, and $A_\mathrm{eff}=85\ \mu\mathrm{m}^2$ for the
effective area of the fibre mode (as per the manufacturer specifications).

\subparagraph*{\hskip-10pt Modelling of cavity solitons.}

Our experimental observations of acoustic-mediated interactions of temporal cavity solitons are modelled with a
mean-field Lugiato-Lefever equation \cite{lugiato_spatial_1987} extended to take into account the electrostrictive
response \cite{dianov_long-range_1992}. Specifically, the evolution of the intracavity electric field $E(t,\tau)$
with respect to the slow-time~$t$ of the resonator is given by a single partial differential equation which reads:
\begin{multline}
  \frac{\partial E(t,\tau)}{\partial t} = \left[ -1-i\Delta-i\eta\frac{\partial^2}{\partial\tau^2}\right]E + S + V \frac{\partial E}{\partial\tau}\\
    + i\left[(1-f)|E|^2 + f \int_0^{+\infty} h(\tau')|E(t,\tau-\tau')|^2 d\tau'\right]E\,.
  \label{eq:LL}
\end{multline}
The normalization of this equation is the same as that used in the Supplementary Information of
Ref.~\citenum{leo_temporal_2010}. The first four terms of the right-hand side of the equation represent,
respectively, the total cavity losses, the detuning of the pump from resonance (with $\Delta$ the detuning
parameter), second-order chromatic dispersion (with $\eta=-1$ the sign of the group-velocity dispersion coefficient
$\beta_2$ of the fibre), and the external driving (with $S$ the amplitude of the driving field). The 2nd line of the
right hand side accounts for the nonlinearity, which is split into an instantaneous (electronic) Kerr contribution
and a delayed response due to the electrostriction-induced acoustic wave. $h(\tau)$ is the acoustic impulse response
function [$h(\tau<0)=0$] normalized such that $\int_0^{+\infty} h(\tau)d\tau = 1$ and $f$ is the fraction of the
Kerr nonlinearity due to electrostriction. With our parameters, we found $f=15.6$\,\% in good agreement with
experimental studies \cite{buckland_electrostrictive_1996}. With these notations, it must be clear that the
nonlinear coefficient~$\gamma$ that appears in the normalization of the field amplitude (see
Ref.~\citenum{leo_temporal_2010}) must be understood as being the nonlinearity coefficient as seen by a c.w.\ wave,
i.e.\ inclusive of the electrostriction contribution \cite{buckland_electrostrictive_1996, agrawal_nonlinear_2006}.
The impulse response of the actual refractive index perturbation~$\delta n$ that is plotted in
Fig.~\ref{fig:acoustic}a was obtained by considering that the intensity of the CS is a Dirac-$\delta$ function. This
leads (in real units) to $\delta n = \gamma f h(\tau) E/(2\pi/\lambda_0)$, where $E=23.3$~pJ is the energy of our CS
(excluding the background) and $\lambda_0=1550$~nm is the driving laser wavelength.

The temporal CSs are the stationary solutions ($\partial E/\partial t=0$) of Eq.~(\ref{eq:LL})
\cite{tlidi_localized_1994}. We have calculated them by looking for the roots of the right-hand side of the equation
with a multi-dimensional globally-convergent Newton-Raphson method. The use of a Newton solver is necessary because
the large difference of scales between the duration of the CSs (a few picoseconds) and the timescale of the acoustic
response (up to tens of nanoseconds) makes direct propagation simulations very inefficient. In presence of the
acoustic response, however, the CSs are not stationary in the reference frame of the driving field. In fact, the CSs
suffer from a self-frequency shift due to the leading edge of the refractive index acoustic perturbation they
generate \cite{dianov_long-range_1992}, resulting in a change in their group-velocity. To circumvent this issue, we
have included in Eq.~(\ref{eq:LL}) a drift term $V\partial E/\partial\tau$ so that the fast-time $\tau$ used to
describe the temporal envelope of the CSs is defined in a reference frame that travels at a velocity~$V$ with
respect to the driving field. The drift velocity~$V$ is treated as an additional unknown in the Newton solver, with
the extra condition that the sought solution must peak in the centre of the computed temporal window. With our
experimental parameters (corresponding to $X=S^2=3.27$ and $\Delta=2.89$, determined as in
Ref.~\citenum{leo_temporal_2010}), we find (in real units) $V=V_1=5.2 \times 10^{-6}$~ps/round-trip for the leading
CS.

In a pair of interacting CSs, the trailing CS moves with respect to the leading one, and the overall situation is
not stationary (except for particular separations). To calculate the velocity of the trailing CS at arbitrary delays
with the Newton method, we took advantage of the fact that the transverse acoustic wave generated by a CS only
affects trailing CSs and not the other way around. This leads to the following approach. First we solve
Eq.~(\ref{eq:LL}) for the leading CS all by itself. Because of the generated acoustic wave, we must note that the
background trailing that CS is weakly perturbed from the normal c.w.\ state. In a second step, the equation is
solved for the trailing CS, again all by itself in the numerical window, but with an important modification: In the
convolution term $\int_0^{+\infty} h(\tau') |E(\tau-\tau')|^2 d\tau'$, the intensity $|E(\tau)|^2$ at the front of
the trailing CS is substituted for the intensity profile of the leading CS and its perturbed trailing background,
with the delay under consideration. In this way, we ensure that the trailing CS sits in the correct perturbed
refractive index wake and the velocity~$V$ output by the Newton solver for the trailing CS includes contributions
both from the self-frequency shift and from the presence of the leading CS. Note that upon convergence, we find no
discontinuity between the field of the trailing CS and the trailing background of the leading CS, therefore
validating our approach. This procedure is repeated for a range of delays (note that the leading CS is only
calculated once). The simulated trajectories shown in Fig.~\ref{fig:20ns}b are then obtained by simple integration,
\begin{equation}
  t(\Delta\tau) = \int_{\Delta\tau_0}^{\Delta\tau} \frac{1}{V(\Delta\tau')-V_1} d\Delta\tau'
\end{equation}
where $V(\Delta\tau)$ is the drift velocity of the trailing CS as a function of its separation $\Delta\tau$ with the
leading one, $\Delta\tau_0$ is the initial separation between the two CSs, and $V_1$ the drift velocity of the
leading CS. Finally, we have verified our modelling approach using direct split-step Fourier integration
\cite{agrawal_nonlinear_2006} of Eq.~(\ref{eq:LL}) with smaller CS separations for which split-step integration does
not impose an unmanageable computational load.

\bigskip

\section*{Acknowledgements}

\noindent This work was supported by the Marsden Fund Council from Government funding, administered by the Royal
Society of New Zealand. The driving laser was funded from the Faculty Research Development Fund of The Faculty of
Science of The University of Auckland. J.K.J. also acknowledges the support of a University of Auckland Doctoral
Scholarship.

\section*{Author Contributions}

\noindent J.K.J. performed the experiments. M.E. identified the physical origin of the interactions, incorporated
the acoustic response into the mean field model, wrote a first draft of the paper, and together with S.C. performed
the numerical simulations. S.G.M. supervised the experimental work. S.C. programmed the Newton solver, wrote the
final version of the paper, and supervised the overall project.

\section*{Additional information}

\noindent The authors declare no competing financial interests. Correspondence and requests for materials should be
addressed to J.K.J. or S.C.

\end{document}